\documentclass[twocolumn]{aa}			
\usepackage{graphicx}

\usepackage{txfonts}
\usepackage{natbib}
\bibpunct{(}{)}{;}{a}{}{,}
\def\gapprox{{_>\atop{^\sim}}}

\def\cmmd{\rm {cm^{-3}}}

\def\s-1{\rm {s^{-1}}}
\def\twco{$^{12}$CO}

\def\HC3N{HC$_3$N}

\def\kms{\hbox{${\rm km\,s}^{-1}$}}
\def\msun{M$_{\odot}$}

\usepackage{url}

\begin{document}
   \title{Detection of HCN, HCO$^+$ and HNC in the Mrk~231 molecular outflow\thanks{Based on observations
carried out with the IRAM Plateau de Bure Interferometer. IRAM is
supported by INSU/CNRS (France), MPG (Germany) and IGN (Spain)}}

   \subtitle{Dense molecular gas in the AGN wind}

   \author{S. Aalto
          \inst{1}
	  \and
	  S. Garcia-Burillo\inst{2}
          \and
          S. Muller\inst{1} 
          \and
          J. M. Winters\inst{3}
          \and
          P. van der Werf\inst{4}
          \and
          C. Henkel\inst{5}
          \and
	  F. Costagliola\inst{1}
          \and
	  R. Neri\inst{3}
             }

 \institute{Department of Earth and Space Sciences, Chalmers University of Technology, Onsala Observatory,
              SE-439 94 Onsala, Sweden\\
              \email{saalto@chalmers.se}
       \and Observatorio Astron\'omico Nacional (OAN)-Observatorio de Madrid, Alfonso XII 3, 28014-Madrid, Spain
       \and Institut de Radio Astronomie Millim\'etrique (IRAM), 300 rue de la Piscine, Domaine Universitaire de Grenoble,
38406 St. Martin d$ ' $H\`eres, France
       \and Leiden Observatory, Leiden University, 2300 RA, Leiden, The Netherlands
       \and Max-Planck-Institut f{\"u}r Radioastronomie, Auf dem H{\"u}gel 69, 53121 Bonn, Germany
	  }

   \date{Received xx; accepted xx}

 
  \abstract
   {}
   {Our goal is to study gas properties in large-scale molecular outflows and winds from active galactic nuclei (AGNs) and starburst galaxies.}
   {We obtained high-resolution (1.\arcsec 55 $\times$  1.\arcsec 28) observations of HCN, HCO$^+$,
HNC 1--0 and HC$_3$N 10--9 of the ultraluminous 
   galaxy (ULIRG)  Mrk~231 with the IRAM Plateau de Bure Interferometer.}
   {We detect luminous emission from HCN, HCO$^+$ and HNC 1--0 in the QSO ULIRG Mrk~231. All three lines show broad line wings - which are
 particularly prominent for HCN. Velocities are found to be similar ($\approx \pm 750$ \kms) to those found for CO 1--0. This is the first time bright
HCN, HCO$^+$ and HNC emission has been detected in a large-scale galactic outflow.  We find that both the blue- and red-shifted line wings
are spatially extended by at least 0.\arcsec 75 ($>$700 pc) in a north-south direction. 
The line wings are brighter (relative to the line center intensity) in HCN than in CO 1--0 and line ratios suggest that the molecular
outflow consists of dense ($n>10^4$ $\cmmd$) and clumpy gas with a high HCN abundance $X$(HCN)$>10^{-8}$.  These properties are
consistent with the molecular gas being compressed and fragmented by shocks in the outflow.
Alternatively, HCN is instead pumped by mid-IR continuum, but we propose that this effect is not strong for the spatially extended outflowing gas.
In addition, we find  that the rotation of the main disk, in east-west direction, is also evident in the HCN, HCO$^+$ and HNC  line emission.
An unexpectedly bright HC$_3$N 10--9 line is detected inside the central 400~pc of Mrk~231. This HC$_3$N emission
may emerge from a shielded, dust-enshrouded region within the inner 40-50~pc where the gas is heated to high temperatures (200 - 300 K) by the AGN.
 }
   {}

   \maketitle
%

\section{Introduction}

Mergers and interactions between galaxies trigger massive starbursts and feed the
growth of supermassive black holes (SMBH) in their centers. Powerful outflows have been proposed
to regulate both star-formation and the SMBH growth through negative feedback in young galaxies \citep[e.g.][]{hopkins09,murray05}. 
\citet{murray05} suggest that momentum-driven winds and outflows are the underlying mechanism
behind the Faber-Jackson relation for large elliptical galaxies. 

In the past years, few studies have targeted outflowing molecular gas \citep[e.g.][]{nakai87,garcia-burillo01,walter02,sakamoto06}
but recently evidence is mounting for massive molecular outflows in AGNs and starburst galaxies \citep{sakamoto10,feruglio10,chung11,alatalo11,sturm11}
and high-velocity molecular outflows have been found in OH-megamasers \citep[e.g.][]{baan}.
It is important to study outflowing molecular gas because this process removes gas immediately available for star
formation from the galaxy which impacts the evolution of the galaxy on short time scales. Furthermore, the molecular gas
properties  hold important clues to the nature and evolution of the outflow. Simulations of molecular cloud properties in galactic outflows have been carried out by e.g. \citet{narayanan}.

The ultraluminous infrared galaxy (ULIRG, log($L_{\rm IR})=12.37$) Mrk~231
is often referred to as the closest infrared quasar (QSO). It is a major merger and hosts powerful AGN activity
as well as a young, dusty starburst with an extreme star-formation rate of $\approx$ 200 \msun
yr$^{-1}$ \citep{taylor99,gallagher02,lipari09}. Massive amounts of molecular
gas reside in an east-west rotating disk \citep{bryant,downes98}.
Radio continuum observations show jets and outflows ranging from pc to kpc-scale \citep{carilli98,lonsdale03}.
\citet{rupke11} discuss the various outflows of Mrk~231 and using high resolution spectroscopy, they
describe a wide, kpc-scale, high-velocity ($\approx$ 1000 \kms) outflow seen in neutral gas (Na~I~D) absorption.
Furthermore, they find that north of the nucleus the radio continuum jet is accelerating the outflow to even higher 
velocities ($\approx$ 1400 \kms) and 3.5~kpc south of the nucleus a slower, starburst-driven wind is found. 
They suggest that the high mass outflow rate of Mrk~231 (which is 2.5 times greater than the star-formation rate)
is a result of the negative feedback from the AGN. Interestingly, the outflow has been detected in molecular gas manifested in the
broad (750 \kms) wings in the CO 1--0 line, as found by \citet{feruglio10}. These authors estimate the outflow of molecular
gas to 700 \msun \,yr$^{-1}$ which could empty the reservoir of molecular gas within 10~Myr.
\citet{feruglio10} argue that the kinetic energy from the supernovae induced by the circumnuclear starburst disk
is not sufficient to expel the gas and that action from the AGN is required. Furthermore, \citet{murray05}  find that the
mass loss rate in a starburst-driven outflow does not significantly exceed the star-formation rate (following their equation (13)). 
The high mass outflow rates of \citet{feruglio10} and \citet{rupke11} seem to support the notion of an AGN-driven outflow. 
This conclusion is disputed by \citet{chung11}, who find  high-speed (1000 - 2000 \kms) molecular outflows in ULIRGs and
argue that they can be powered by starbursts.
They suggest that the mass outflow rate of Mrk~231 is overestimated if the CO emission in the outflow is optically thin.
Clearly, a more detailed study of the properties of the gas in the outflow is essential for a better estimate of the mass
in the outflow, but also to understand how the molecular gas is carried out of the galaxy and by what process. Is the molecular gas entrained in outflowing hot material - or is it being expelled by radiation pressure from an AGN and/or the momentum flux by supernovae?
How can the the gas survive in molecular form - and for how long?

In our IRAM 30m 3~mm EMIR spectrum \citep{costagliola11} of Mrk~231 we found that the HCN and HCO$^+$  
line widths were surprisingly broad, leading us to suspect that the outflow signature of Mrk~231 could be even more powerful in the 
rotational transitions of these high-density tracers. To confirm this discovery
and to map the extent of the line emission, we asked for IRAM Plateau de Bure DDT time to observe HCN, HCO$^+$ and HNC in the
B-array.  

In sections~\ref{s:obs} and \ref{s:res} the observations and their results - bright HCN, HCO$^+$ and HNC 1--0 emission
in the outflow line wings and a serendipitous detection of the HC$_3$N 10--9 line - are described.
In sections~\ref{s:res1} and \ref{s:res1} we discuss the spatial extent of the line wings, and in sections~\ref{s:wings} and \ref{s:hcn} potential underlying causes for the enhanced wing emission are outlined.  Outflow gas properties and the nature of the outflow 
are discussed in section \ref{s:properties}. In section~\ref{s:hc3n} we discuss the origin
of the HC$_3$N line emission, and finally in section~\ref{s:future} a brief future outlook is outlined.


\begin{figure*}
\resizebox{18cm}{!}{\includegraphics[angle=0]{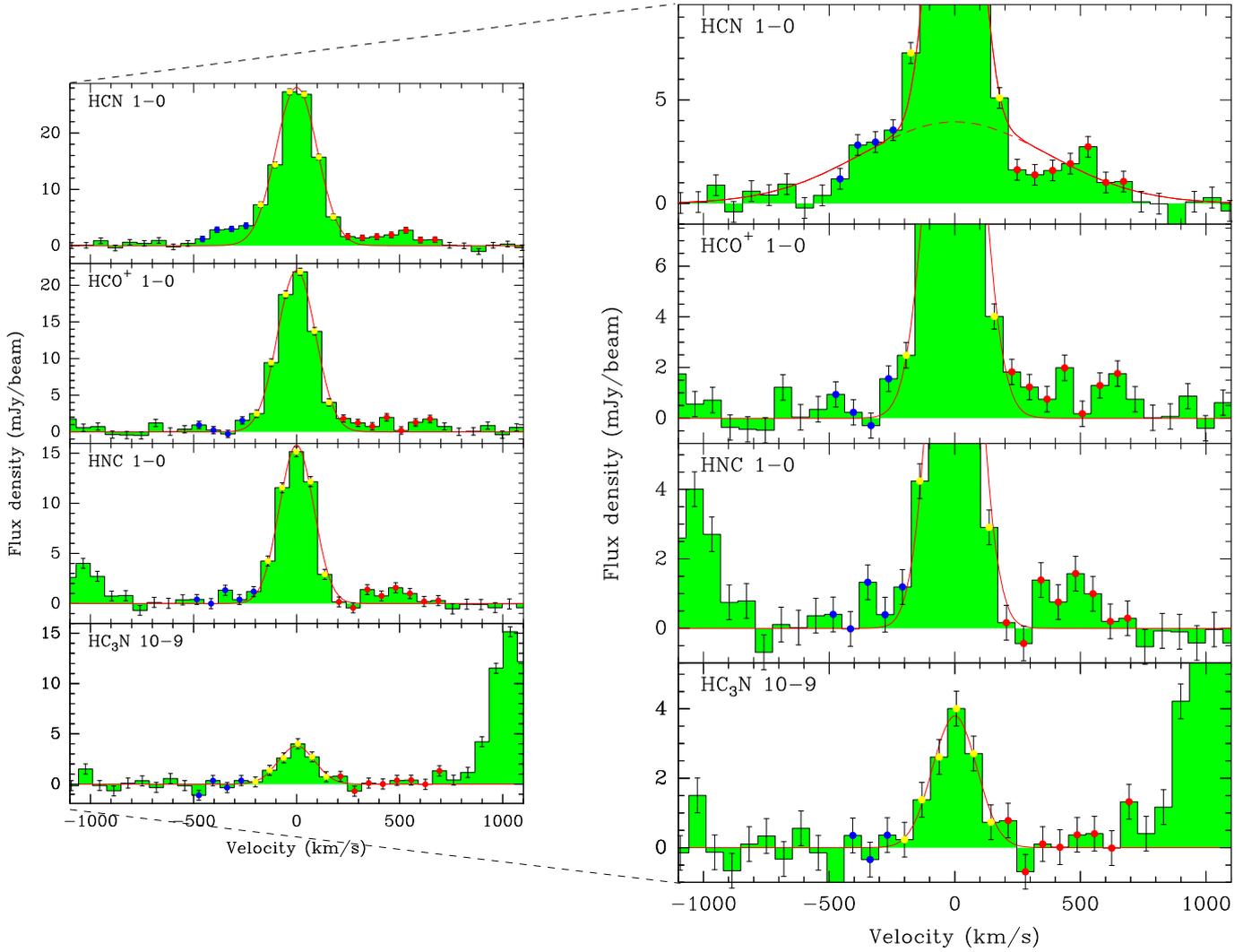}}
\caption{\label{f:spec} Plateau de Bure spectra of HCN, HCO$^+$, HNC 1--0 and HC$_3$N 10--9.
In the right panels we have zoomed-in on the base of the line to show the line wings more clearly. 
In general, red solid lines show Gaussian fits to the line center (core) line widths and are 197~\kms for HC$_3$N and HNC,
213 \kms for HCO$^+$ and 236~\kms for HCN. However, in the top right panel of HCN 1--0 we instead show two Gaussian fits 
(one with $\delta V$(FWHM)=180 \kms\ and one with $\delta V$(FWHM)=870 \kms), as discussed in section~\ref{s:res1}.
Zero velocity was set for redshift $z$=0.042170.}
\end{figure*}


\section{Observations}
\label{s:obs}

The observations were carried out with the six-element IRAM Plateau de Bure array on 
March 8, 2011, in B-array.  The phase center was set to $\alpha$=12:56:14.232 and $\delta$= 56:52:25.207 (J2000).
Mrk~231 was observed for hour angles 2 (i.e. starting 2 hrs after source culmination) to 5 and the radio seeing was $0.\arcsec 8$.
The receivers were tuned to a frequency of 85.3125~GHz, the center between the two red-shifted line frequencies  for HCN ($\nu$=85.046~GHz)
and HCO$^+$ ($\nu$=85.579~GHz). Zero velocity refers to the redshift $z$=0.042170 \citep{carilli98} throughout the paper,
resulting in a spatial scale of 1\arcsec=870~pc. We used the WideX correlator, which provides a broad frequency range of
3.6~GHz, which also included the HNC 1--0 and HC$_3$N 10--9 lines.

The bandpass of the individual antennas was derived from the bright quasar 3C273.
The flux calibration was set on 3C273. The close-by quasars $J1259+516$ ($\sim 0.32$ Jy at 3 mm) and
$J1300+580$ ($\sim 0.19$ Jy) were observed regularly every 25 minutes for complex gain calibration.

After calibration within the GILDAS reduction package, the visibility
set was converted into FITS format, and imported into the GILDAS/MAPPING and AIPS packages
for additional imaging. Data were rebinned to 20~MHz (68 \kms) and cleaned with a robust weighting of 3, 
resulting in a beam size of 1.\arcsec55$\times$1.\arcsec28 and position angle PA=57$^{\circ}$. 
The corresponding rms noise is 0.48~mJy~channel$^{-1}$.


\section{Results}
\label{s:res}

\subsection{HCN }
\label{s:res1}

Spectra of all four detected lines are presented in Fig.~\ref{f:spec}. Prominent line wings are detected in the HCN 1--0
line out to velocities  of 750 \kms, similar to those found in CO 1--0 by \citet{feruglio10}. 
If we define the line-center-to-wings ratio (LCW) as the ratio between peak intensity at zero velocity to the intensity at +500 \kms in
the red-shifted line wings, then the LCW for HCN 1--0 is 27~mJy/3~mJy = 9. In the CO 1--0 spectrum of \citet{feruglio10}
the LCW is 320~mJy/8~mJy = 40. Another way of defining the LCW is to adopt the Gaussian fits in Fig.~1 of \citet{feruglio10},
where two components, one with $\delta V$(FWHM)=180 \kms\ and one with $\delta V$(FWHM)=870 \kms, are fitted to the line center
and outflow respectively. Fixing the linewidths of the two Gaussian componets to our data
(see Fig.~\ref{f:mom0} for the Gaussian fit), we find an LCW for HCN of 7.7 and for CO 1--0 we estimate an LCW of 20.6.

Because the line wings are most prominent for HCN, we elected to show the line integrated intensity of the wings for this molecule only. The HCN 1--0 integrated intensity maps of the line wings and the total line emission are presented in Fig.~\ref{f:mom0}, peak- and
integrated intensities in Tab.~\ref{t:flux}. The total line emission is beginning to be resolved in the  1.\arcsec55$\times$1.\arcsec28  beam with features extending out to the west and northeast. These are generally consistent with similar features in the CO 2--1 map by \citet{downes98} and
in the HCO$^+$ 4--3 map by \citet{wilson08}. The central velocity gradient from the main disk is evident in HCN 
- showing the $\pm$ 150 \kms  gradient seen in CO \citep{downes98}. The HCN 1--0 velocity field of the central component
(wings excluded) is presented in Fig.~\ref{f:mom1}. HCN position velocity diagrams along the rotational axis (PA=90$^{\circ}$) and
along the wings (PA=0$^{\circ}$) are presented in Fig.~\ref{f:pv}.

\subsubsection{Position of line wings and continuum subtraction}
\label{s:sub}

\begin{figure}
\resizebox{7cm}{!}{\includegraphics[angle=0]{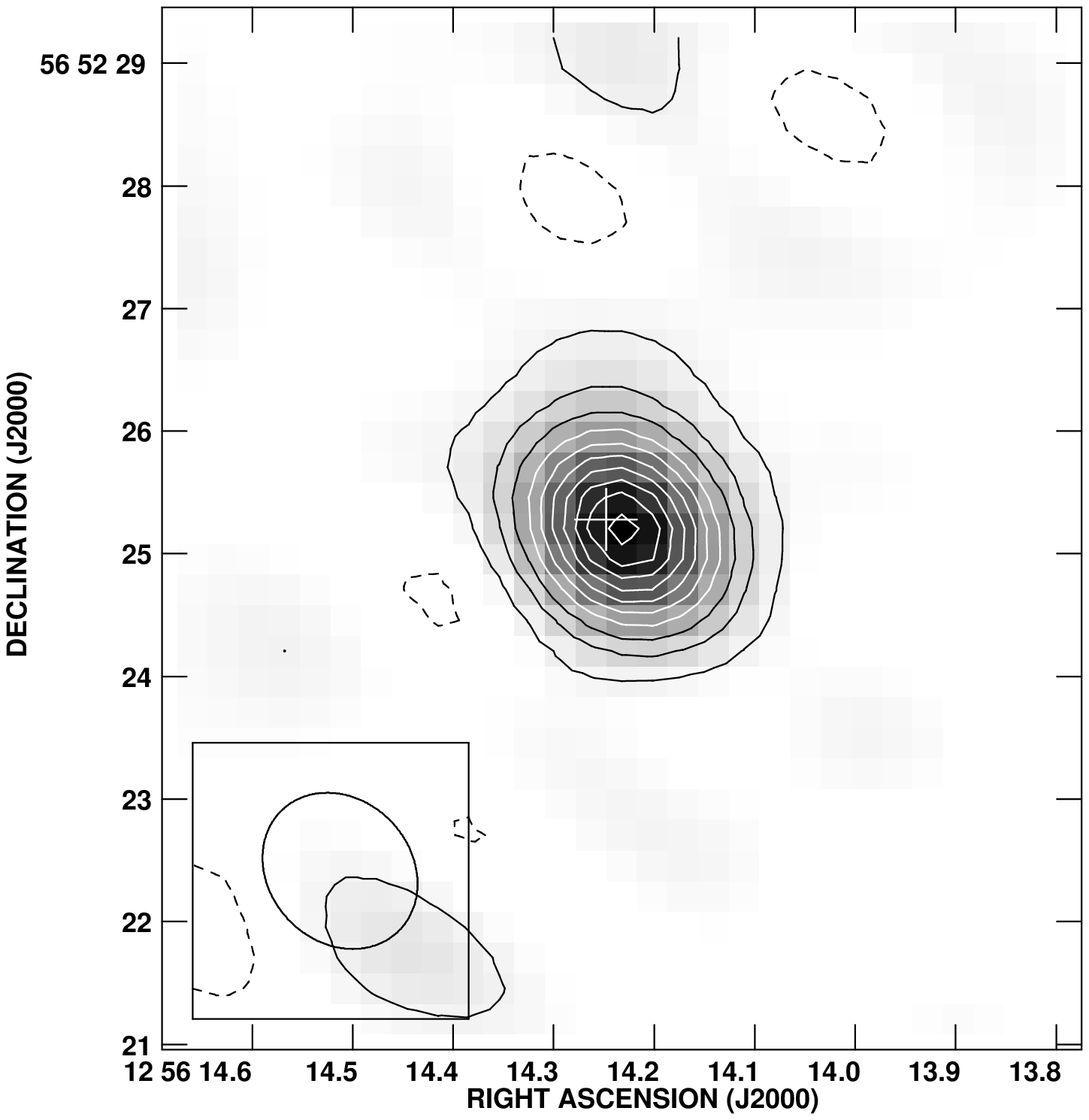}}
\caption{\label{f:cont} 85~GHz continuum image of Mrk~231, the VLBI position is marked with a cross
(see footnote to Tab.~\ref{t:flux}).  }
\end{figure}

Because the line wings are broad and relatively faint, the determination of their positions depends on a reliable continuum subtraction.
We used two methods of subtracting the continuum to test the relative errors in the wing position: subtraction in
the uv-plane and in the image plane. The continuum was subtracted with a zeroth or first order fit to the line-free
spectral channels. In general we find that the uncertainty in the wing peak position is 0.\arcsec 4. The continuum map is presented in
Fig.~\ref{f:cont}. A Gaussian fit to the continuum
source gives a flux density of $25.0 \pm  0.6$ mJy and an upper limit to the source size of 0.\arcsec 39.

\begin{figure*}
\resizebox{18cm}{!}{\includegraphics[angle=0]{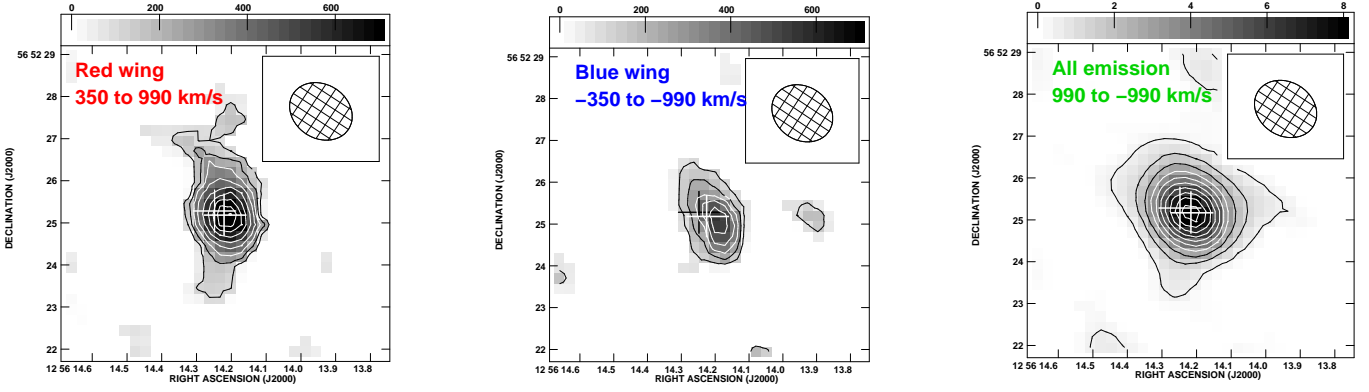}}
\caption{\label{f:mom0} Integrated intensities of the HCN 1--0 line emission. Left panel: The red-shifted wing emission from
350 to 990~\kms. The grayscale range is from 0 to 0.7 Jy~\kms~(beam )$^{-1}$ and the contours start at 0.098 Jy~\kms~(beam)$^{-1}$ with steps of 0.0998 Jy~\kms(beam)$^{-1}$. Middle panel: The blue-shifted wing emission from -990 to -350 \kms. Same grayscale and contours as for the previous panel. Right panel: The total (from -990 \kms to 990 \kms) integrated HCN 1--0 line
emission from Mrk~231. The grayscale range is from 0 to 8 Jy~\kms~(beam)$^{-1}$ and the contour levels are
0.45 Jy~\kms~(beam)$^{-1}\times$(1,3,5,7,9,11,13,15,17,19). The thick white crosses mark the peak of the
integrated HCN 1--0 line emission and the thinner crosses indicate the VLBI position (see footnote to Tab.~\ref{t:flux})
and 1\arcsec=870~pc.}
\end{figure*}

\begin{figure}
\resizebox{8cm}{!}{\includegraphics[angle=0]{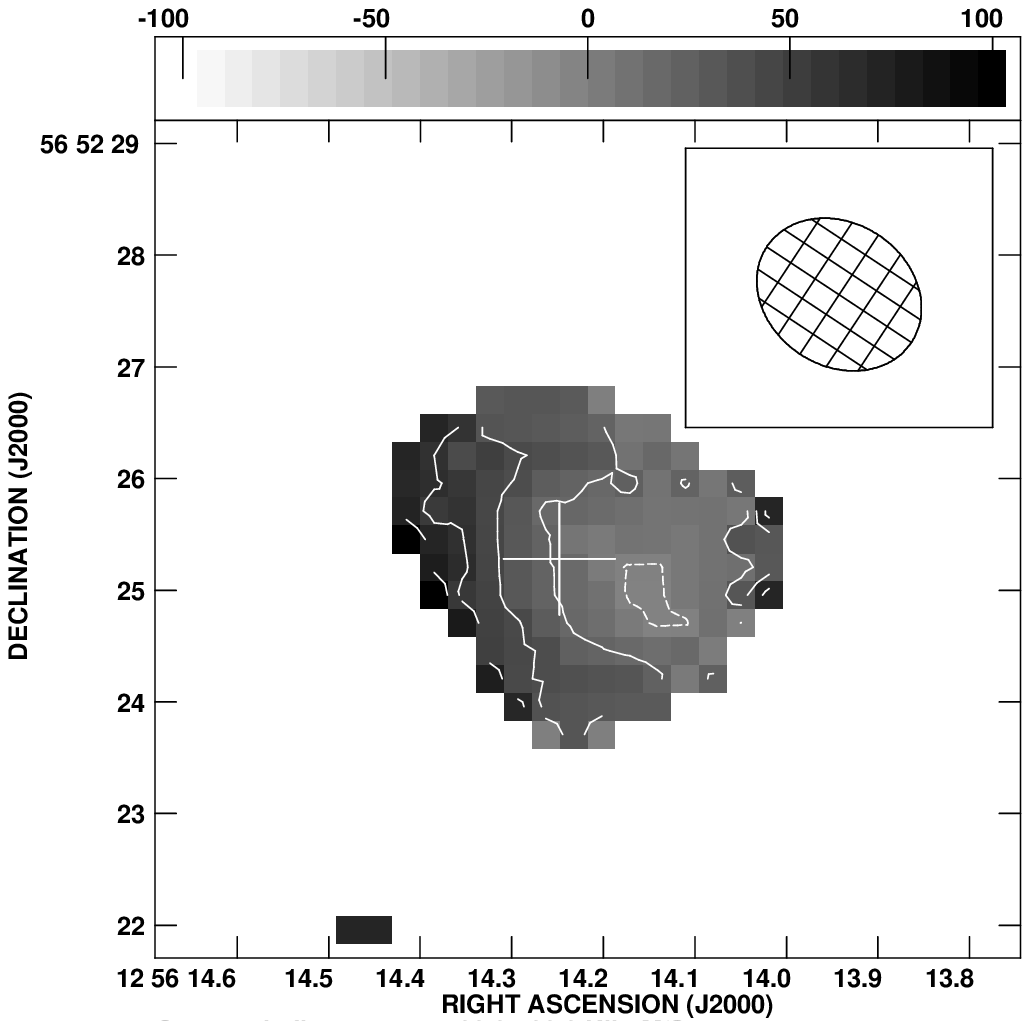}}
\caption{\label{f:mom1} HCN 1--0 velocity field (including only velocities $\pm$200 \kms)
showing the rotation of the molecular disk. The HCN 1--0 emission rotates in the same east-west fashion as the \twco\ emission
\citep{bryant,downes98}. The grayscale ranges from -100 to 100 \kms and the contours are 20~\kms~$\times$ (-4,-3,-2,-1,0,1,2,3,4).
Zero velocity was set to redshift $z$=0.042170 and 1\arcsec=870~pc. Cross marks the VLBI radio continuum position
(see footnote to Tab.~\ref{t:flux}).}
\end{figure}

\begin{figure*}
\resizebox{18cm}{!}{\includegraphics[angle=0]{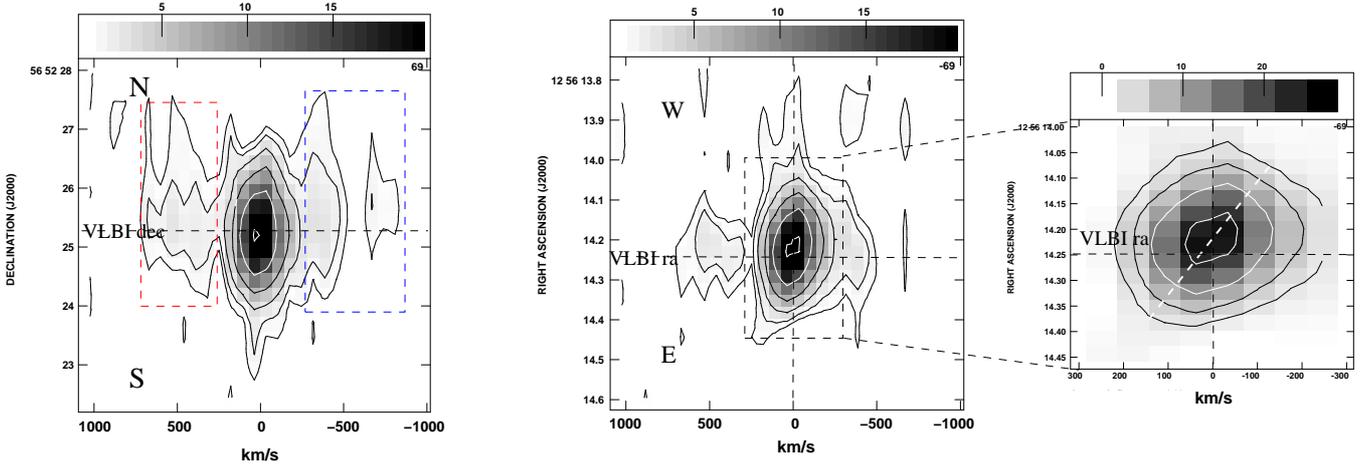}}
\caption{\label{f:pv} HCN 1--0 position-velocity (pV) diagrams. Left: North-south cut along the minor axis of the disk rotation. 
The extended nature of the line wings is evident here.
Center: East-west cut (along the major axis of the rotating disk) and Right: Zoom-in on the disk rotation.(1\arcsec=870~pc) }
\end{figure*}

The blue- and red-shifted HCN line-wings appear shifted by 0.\arcsec2 -- 0.\arcsec4 (175 -- 350 pc) to the southwest
of the nucleus as defined by the peak position of the integrated HCN 1--0 emission (see Fig.~\ref{f:mom0}). 
Given the positional uncertainties, we may conclude that the wings have their peak brightness within 0.5~kpc of
the nucleus. Furthermore, a faint blue-shifted wing feature can be seen tentatively 2\arcsec to
the west of the nucleus (a similar feature is seen around the systemic velocity). Faint red-shifted wing emission
is curving out 2\arcsec  northeast of the center. These features should be confirmed at higher sensitivity.


\begin{table}
\caption{\label{t:flux} Line flux densities$^a$.}
\begin{tabular}{lccc}
 & \\
Line & Peak & Wing$^b$ & Integrated  \\
\hline
& [mJy\,beam$^{-1}$] & [mJy\,beam$^{-1}$] & [Jy \kms]\\
\hline \\ 
HCN 1--0 & 28 $\pm$ 0.5 & 3 & 8.4 \\
HCO$^+$ 1--0 & 22 $\pm$ 0.5 & 2 & 5.6 \\ 
HNC 1--0 & 15 $\pm$ 0.5 & 1 & 3.7 \\
HC$_3$N & 3 $\pm$ 0.5 & $\dots$ & 0.9 \\
\\
\hline \\
\end{tabular} 

a) The 3mm continuum position is $\alpha$: 12:56:14.230 $\pm$ 0.0013 \, 
$\delta$: 56:52:25.206 $\pm$  0.01 (J2000). A Gaussian fit to the continuum
source gives a flux density of $25.0 \pm  0.6$ mJy and an upper limit to the source size of 0.\arcsec 39.
The VLBI radio continuum peak is: $\alpha$: 12:56:14.248 \, $\delta$: 56:52:25.28 (J2000) \citep{carilli98}. 

b) From the red-shifted wing at $V$=+500 \kms.

\end{table}


\subsection{HCO$^+$, HNC and HC$_3$N}
\label{s:res2}

\begin{figure*}
\resizebox{16cm}{!}{\includegraphics[angle=0]{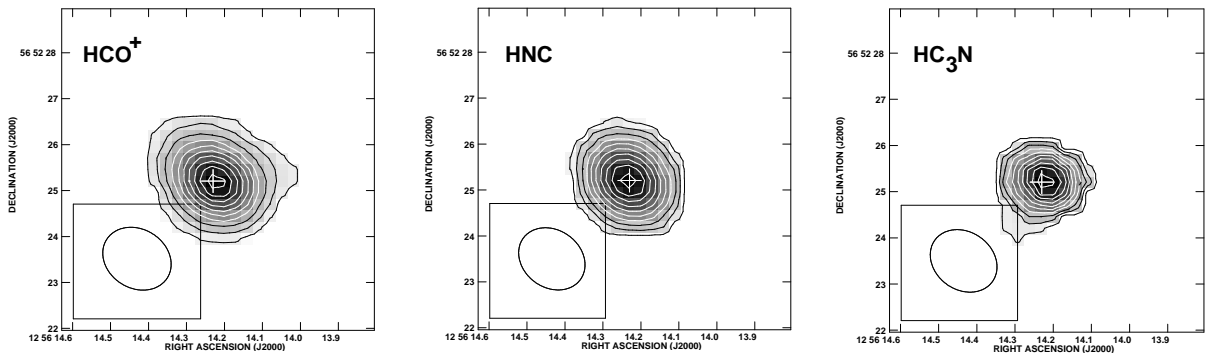}}
\caption{\label{f:other_mom0} Integrated intensities (from left to right) of the HCO$^+$, HNC 1--0 and HC$_3$N 10--9 line emission. 
The grayscale range is from 0 to 5.3, 0 to 3.3, and 0 to 1.0 Jy~\kms~(beam)$^{-1}$ respectively. 
The contour levels are 0.26, 0.16, and 0.04 Jy~\kms~(beam)$^{-1} \times$(1,3,5,7,9,11,13,15,17,19).
(1\arcsec=870~pc and and zero velocity was set to redshift $z$=0.042170 for all lines). Cross refers to the center of the 85~GHz continuum. The maps are 
integrated over $\pm$200 \kms and exclude the line wings. }
\end{figure*}

\begin{figure}
\resizebox{7cm}{!}{\includegraphics[angle=0]{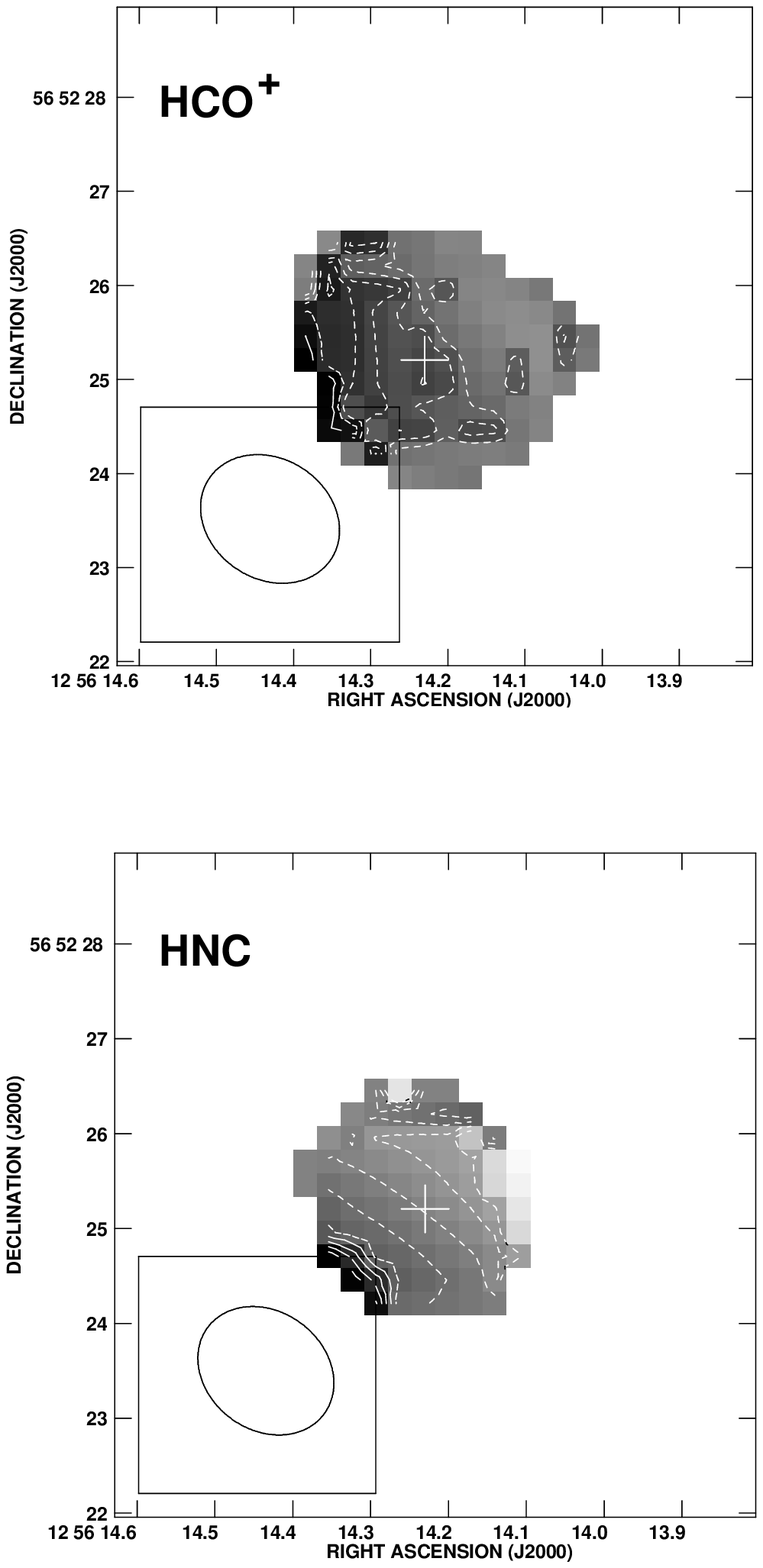}}
\caption{\label{f:other_mom1} Central HCO$^+$ (top) and HNC (bottom) 1--0 velocity fields. 
The grayscale range is from -64 to 9 \kms (top) and from -43 to 24 \kms (bottom).
The contour levels are 5 \kms $\times$(-4,-3,-2,-1,0,1,2,3,4).
(1\arcsec=870~pc and zero velocity was set to redshift $z$=0.042170 for all lines). Cross refers to the center of the 85~GHz continuum. The emission is integrated over $\pm$200 \kms and excludes the line wings. 
}
\end{figure}

In the HCO$^+$ and HNC spectra (see Fig.~\ref{f:spec} and Tab.~\ref{t:flux}) the wings can also be seen - although they are
weaker than for HCN. The peak ($V$=0 \kms) line intensity ratio between HCN and HCO$^+$ is 1.3, while the 
ratio in the wings (summed from +200 to +700~\kms and from -200 to -700~\kms) is $\approx$2.2.  For comparison, the
integrated line ratio is reported to be $\approx$1.8 in the single-dish survey of \citet{costagliola11}.  The peak HCN/HNC line intensity ratio is 1.8 and the line wing ratio is $\approx$2.8. For comparison, the integrated line HCN/HNC ratio is $\approx$2.6 in
the single-dish survey of \citet{costagliola11}. The rotation of the main disk is evident for both HCO$^+$ and HNC (see Fig.~\ref{f:other_mom1}). The  line center integrated intensity maps of HCO$^+$ and HNC are presented in Fig.~\ref{f:other_mom0}.

Emission from the HC$_3$N molecule is found for the first time in Mrk~231 through the detection of the  HC$_3$N $J$=10--9 line
(see Fig.~\ref{f:other_mom0}). Its line width is similar ($\approx 197$ \kms) to that of the HNC line center.
The HCN 1--0/HC$_3$N 10--9 peak line ratio is 7.7 (in the \citet{costagliola11} single-dish survey a lower limit of 14 is found).
The HC$_3$N 10--9 emission is unresolved in the 1.\arcsec55$\times$1.\arcsec28 beam.


\section{Discussion}
\label{s:dis}

\subsection{The extent of the HCN line wings}
\label{s:wings}

We confirm the result of \citet{feruglio10} that the line wings are emerging from a spatially extended
structure. This supports the notion that they originate in an extended outflow - rather than representing 
the extremes of a Keplerian nuclear disk. (The notion of extended Keplerian rotation fails on the extreme required enclosed
mass.) A two-dimensional Gaussian fit to the integrated intensity map of the red-shifted wing results in a 
full-width-at-half-maximum (FWHM) size of 0.\arcsec85 $\pm$ 0.\arcsec2 (1\arcsec=870~pc).
The corresponding map for the blue-shifted wing has lower signal-to-noise and a Gaussian fits results in
a FWHM size of 0.\arcsec75 $\pm$ 0.\arcsec3. 

Comparing the HCN and HCO$^+$ wings with the structure of the atomic (neutral) gas as seen in absorption
\citep{rupke11}, the results are consistent with an almost face-on nuclear wind.  
\citet{rupke11} find the most highly accelerated atomic gas in the northern region (with peak velocities occurring around 2.\arcsec3
(2~kpc) north of the nucleus) where the radio jet is interacting
with the neutral material. \citet{rupke11} suggest that the jet is entraining material from the large-scale wind that 
accelerates it even more which is a phenomenon seen also in other AGNs \citep{matsushita07}.
There is a hint of this feature at the 1.5 -2$\sigma$ level in our HCN integrated intensity map of the red-shifted wind - but this requires confirmation at higher sensitivity.
Furthermore, the velocities of the nuclear wind described by \citet{rupke11} 
and the HCN wings generally agree well. The molecular HCN wings we find occur equally
close (possibly even closer) to the nucleus as the atomic wind. We find no correspondence to the starburst wind
found 3\arcsec south of the nucleus by \citet{rupke11} and we conclude that the HCN wings originate in the nuclear wind, as does the
atomic wind.

\subsection{Why is the HCN emission enhanced in the line wings?}
\label{s:hcn}

The CO 1--0 wing peak intensity is 8~mJy \citep{feruglio10} and for HCN 1--0 it is 3~mJy. For the same
source size - and transferring to temperature scale - the $T_{\rm B}$(CO)/$T_{\rm B}$(HCN) 1--0 ratio in
the wings is 1.6 - which is lower than for the line peak at center velocity. This is either because the
outflowing molecular gas is dense, with elevated HCN abundances, or it is caused by mid-IR pumping of HCN. 

\subsubsection{Dense molecular gas in the outflow}
\label{s:dense}

If we assume that the HCN molecule is collisionally excited, the density in the outflow has to be at least $n=10^4$ $\cmmd$ for
the line to begin to be excited, even if the HCN abundance is enhanced. 
We can use a simple non-LTE RADEX \citep{vandertak07} model to investigate the CO/HCN 1--0 line ratios in a few possible scenarios.
The models are not unique and serve only as illustrations of possibilities. (Note also that without multi-transition HCN information we are unable to constrain the properties of the dense gas in the outflow.)

We adopt a temperature of the outflowing gas of $T_{\rm k}$=100 K (the lowest gas temperature fit of \citet{gonzalez10} from {\it Herschel} observations of the molecular gas of Mrk~231) and note that for 
densities $n=10^4$ $\cmmd$ solutions require an HCN abundance (relative to H$_2$)  of X(HCN)$\gapprox 10^{-6}$. 
It is actually the relative CO and HCN column densities that are constrained by RADEX - this means that we have to assume a
standard CO to H$_2$ abundance ratio (in this case of 10$^{-4}$) to obtain the HCN abundance X(HCN).
Such an HCN abundance would be considered very high (but see Sect.~\ref{s:hcn_abundance} below for a brief discussion).
For a number density $n=10^5$ $\cmmd$ a line ratio between CO and HCN 1--0 of 2 can be achieved
for optically thin ($\tau < 1$) CO and HCN with a column density ratio ($N$(HCN)/$N$(CO)) of $5 \times 10^{-4}$. If we assume a CO abundance of $10^{-4}$ (relative to H$_2$), this would imply an HCN abundance of X(HCN)=$5 \times 10^{-8}$, which is high - but
not unusually so (see Sect.~\ref{s:hcn_abundance} below). Intrinsic brightness temperatures for optically thin emission would
be low, however, and the emission would have to have a large filling factor. The low HCN/H$^{13}$CN 1--0 intensity ratio of 5.5
observed by \citet{costagliola11} suggests that the  HCN 1--0 emission towards Mrk~231 is not optically thin. This is a global
ratio, however, and the ratio for the wing component is difficult to determine.  

A low intensity ratio between CO and HCN can also be achieved by a clumpy outflow where the individual clumps are dense and
optically thick.  A $T_{\rm B}$(CO)/$T_{\rm B}$(HCN) ratio of $\approx$2  can be obtained for a range of abundances depending
on how high the CO column density can become in each clump. If we avoid the most extreme optical depths in CO, then HCN
abundances of X(HCN)= $10^{-8} - 10^{-7}$ will fit the observed line ratio. This scenario (or the optically thin one above)
does not allow any CO 1--0 emission to emerge from a diffuse, low-density ($n=10^2 - 10^3$  $\cmmd$) component in the outflow
- such a component would only contribute to the CO 1--0 luminosity, not the HCN. 
Therefore, if the molecules are collisionally excited, the low intensity ratio between CO and HCN in the outflow leaves little
room for a low-density molecular component.

\subsubsection{Outflow abundances}
\label{s:hcn_abundance}

\noindent
{\it HCN abundances:} \, Elevated HCN abundances have been observed in outflows in the Galaxy. For example, HCN abundances
of $14 \times 10^{-8}$ are found in the outflow lobes of regions of massive star-formation \citep[e.g.][]{su07,liu11}. 
The fact that the outflow in Mrk~231 is seen (tentatively) in H$_2$O absorption \citep{gonzalez10} may suggest that shocks
are involved in removing the water from the grains. Are we seeing indications of HCN abundances being enhanced in the shocks of the outflow? Evidence that this can be the
case are presented by \citet{tafalla10}. They argue that HCN is mainly produced in shocks by the reaction of N with CH$_2$ - where 
CH$_2$ is formed in the reaction between C and H$_3^+$ or C$^+$ with H$_2$. Thus, for HCN to be selectively produced in the
shock, a sufficient abundance of free C is required. SiO emission from shocks in galaxy-scale outflows has been observed
in the starburst wind of M~82 \citep{garcia-burillo01} and may also be expected to be elevated in the Mrk~231 outflow.

Note also that a large HCN abundance is a general result of high temperatures. \citet{harada10} have shown that HCN abundances
may reach $10^{-6}$ in regions where $T\gapprox 300$~K. They also find enhancements of H$_2$O abundances in their high-temperature models  - therefore HCN and H$_2$O abundance enhancements may be a result of high temperatures and not specifically shock chemistry. A potential scenario is that the molecular abundances reflect the high temperatures near the AGN and retain these abundances in the outflow. However, without a source of heating, the outflowing gas will cool very quickly if it remains at high density and
potential ion-neutral reactions (involving e.g H$^+$, H$_2^+$, O$^+$, H$_3$O$^+$ etc) may destroy HCN on short time-scales. 
So it is unclear whether the gas can retain relic high-temperature abundances of HCN. Furthermore,
at least part of the outflow will be exposed to the X-rays of the nuclear AGN, which means that it could possibly influence the
chemistry of the outflowing gas \citep[e.g.][]{maloney96,meijerink05,meijerink06}.

\noindent
{\it HNC and HCO$^+$ abundances:} \, The HCN intensity is greater than that of HCO$^+$ by a factor of two in the line wings. This could be caused by excitation or by HCN being more abundant than HCO$^+$. In Galactic outflow sources, I(HCN)$>$I(HCO$^+$)
(and abundance ratios HCN/HCO$^+$=10-50) are not uncommon \citep{tafalla10} and may simply be a result of the enhanced HCN abundance
in the outflow due to shocks and high temperatures.  The line wing HCN/HNC ratio of 2.8 is within the range of what is typically
observed in galaxies \citep[e.g.][]{costagliola11}.  In general HNC is expected to be destroyed in shocks - and at high
temperatures \citep[e.g.][]{schilke92}.
For the second model in section~\ref{s:hcn} an abundance ratio HCN/HNC and HCN/HCO$^+$ of $\approx$20 may reproduce the observed 1--0
line ratios in the line wings. Both these abundance ratios are consistent with values found for Galactic shock regions.

\subsubsection{IR pumping}
\label{s:pump}

An alternative scenario to collisional excitation of HCN is pumping by 14 $\mu$m mid-IR continuum. The rotational levels are
then populated via the IR bending modes. Vibrational line emission from HCN has been observed in ULIRGs such as Arp~220 \citep{salter}
and in LIRGs like NGC~4418 \citep{sakamoto10}, showing that the HCN molecule is pumped by mid-IR absorption in these galaxies. For
HCN to be pumped a minimum 14 $\mu$m brightness temperature of $T_{\rm B}$(IR)=85~K is required \citep[e.g.][]{aalto07}. If
IR pumping dominates the HCN excitation no dense clumps in the outflow are needed to explain the HCN emission. It is therefore
important to settle the question of the excitation of HCN in the outflow. Furthermore, both HCO$^+$ and HNC may be
pumped by mid-IR emission \citep{aalto07} through bending modes in the same way as for HCN. HNC is pumped already at  21~$\mu$m and
HCO$^+$ at 12~$\mu$m. 

In the SED for Mrk~231 \citep{gonzalez10} the 14 $\mu$m regime is dominated by a hot component with $T_{\rm d}$=150-400 K emerging
from the inner 40 -- 50~pc. One would expect the strongest pumping of HCN close to this hot component, i.e., the
nucleus. Furthermore, the plane of the disk is likely opaque at 14 $\mu$m, but the gas lifted out of the plane in the outflow
will have an unobstructed view of the nuclear region. Very tentatively, one might argue therefore that the HCN in the wings is pumped
more strongly than that in the core of the line. 
However, the extended nature of the wings implies that the gas is at least a few hundred pc from the IR source and the resulting
brightness temperature produced by this source is going to be well below 85~K, rendering the pumping scenario unlikely.
This argument holds also for HNC and HCO$^+$.

It is interesting to compare the potential HCN pumping scenario to the case for water detected by Herschel \citep{gonzalez10}.
H$_2$O is pumped at significantly longer wavelengths (than HCN) from 45 to 75 $\mu$m and may therefore be pumped by the warm, more
extended dust component ($r$=120~pc disk with $T_{\rm d}$ = 95 K \citep{gonzalez10}). This region is smaller than that of the HCN line wings and 
the H$_2$O absorption line profile also appears different from that of the HCN line wings. It is therefore possible that  H$_2$O is IR-pumped while
HCN is not.


\subsection{Gas properties and the nature of the outflow}
\label{s:properties}

Our interpretation of the HCN-luminous wings of Mrk~231 is that the molecular gas in the outflow is clumpy and dense, with an
enhanced HCN abundance. The similar spectral shapes (and spatial extent within the errors) of the CO and HCN 1--0 line wings
suggest that the emission is emerging from the same outflowing gas. This would then imply that there is no significant evolution (e.g. star-formation or evaporation) of the cloud properties along the molecular part of the outflow. 
This is inconsistent with the simulations of molecular cloud properties entrained in outflows by \citet{narayanan}, who
predict a shorter lifetime and smaller spatial extent of the dense gas phase - as well as ongoing star-formation in the outflowing molecular gas. The Mrk~231 molecular outflow appears instead to be dominated by the dense phase without much low-density
molecular gas. Is this because radiation dissociates the low-density gas, or because it becomes compressed in
the flow? Alternatively, the molecular cloud structure largely reflects the interstellar medium structure from where it becomes
launched into the outflow. Note that the spatial extent of the CO and HCN wings must be better determined for a deeper exploration
of the evolution of the gas in the outflow.

In an AGN-driven outflow, the molecular gas can be entrained by a thermal hot wind or driven by momentum deposition.
\citet{rupke11} find that the neutral gas has higher velocities than the ionized gas in Mrk~231, which is not consistent
with entrainment by a hot wind. They suggest the AGN-driven two-phase model by \citet{hopkins10} as a possible scenario for the neutral wind of Mrk~231:  Close to the center, the AGN acts on the hot interstellar medium, powering an outflow that causes the cold molecular clouds to expand, which makes them more susceptible to being carried outward by radiation pressure (and to become ionized).  However, this requires a mechanism to rapidly compress the molecular gas in the subsequent radiation pressure-driven wind for it to be consistent with our observations. Alternatively, radiation pressure-driven outflows could be launched directly, close to the black hole  \citep[e.g.][]{king03}. Dense molecular gas from the inner regions of Mrk~231 may then be compressed even more in the outflow without being initially disrupted. High HCN abundances may indicate shock enrichment in the outflow and/or that the chemistry is influenced by the AGN.
It is interesting to speculate if HCN-luminous wings may be a general signature of AGN-driven outflows in luminous galaxies. 
High-sensitivity HCN surveys of starburst and AGN-driven outflows will provide us with an answer.

We note that \citet{chung11} detected CN emission only toward the AGN-driven objects, in their stacked survey of outflows in ULIRGs, not toward the starbursts. This is either because of higher gas densities in these outflows, or because of CN abundance enhancements, and also emphasizes the importance of studying the molecular gas properties of outflowing AGN and starburst galaxies.

\subsection{Serendipitous HC$_3$N detection:  warm, shielded gas in the nucleus.}
\label{s:hc3n}

The fitted upper size limit of the HC$_3$N 10--9 integrated line emission is FWHM=0.\arcsec48 (400~pc) and it is centered
on the nucleus of Mrk~231. Emission is not detected in the line wings. The peak-to-peak line ratio between HC$_3$N 10--9 and
HCN 1--0 is 0.14$\pm$0.015 - which makes the HC$_3$N line emission (relative to HCN) brighter than in, for example, NGC~1068 or NGC~7771
- and almost as bright as in Arp~220 \citep{costagliola11}. Models by \citet{harada11}
indicate that HC$_3$N abundances may be enhanced substantially (orders of magnitude) in warm ($>$ 200 K) regions in the midplane of
dense disks around AGNs. These regions do not qualify as XDRs - X-ray dominated regions - but may be buried within the XDR.
According to models based on the recent Herschel observations \citep{vanderwerf10}, the gas in the 1.1~kpc starburst disk is exposed to intense UV emission - which should effectively destroy the HC$_3$N molecules. Since the size of the HC$_3$N emitting region is smaller than the suggested dimensions of the starburst disk, the HC$_3$N emission may emerge from a region between the inner edge of the starburst disk and the
AGN dominated XDR. In the dust SED model  by \citet{gonzalez10} the nuclear hot component with $T_{\rm d}$=150-400 K originates
in the inner 40 -- 50~pc. This may therefore be the location from where the HC$_3$N is emerging - consistent with the expected HC$_3$N abundance enhancement in warm, dusty regions. The high bulk temperature required for significant abundance enhancement is consistent with the AGN as the heating agent.
It is possible that this dusty, warm region represents the final stage of the obscured, X-ray absorbed \citep[e.g.][]{page} accretion phase of the QSO.

\subsection{Future outlook}
\label{s:future}

We have seen that the molecular outflow in Mrk~231 has surprising gas properties and physical conditions that contain important clues
to its origin, physical conditions, and chemical composition. To fully understand the enhancement of HCN (and also the
HCO$^+$ and HNC emission), multi-transition, high-resolution observations are necessary. Other species such as multi-line CO
observations, isotopic variants of CO, HCN, HCO$^+$ and HNC, shock tracers like SiO and CH$_3$OH will provide essential, new pieces to 
this puzzle. These studies should be accompanied by radiative transport and astrochemical modeling. Deep imaging of CO and HCN will
test the notion that the emission is co-spatial and originates in the same gas. 

\noindent
The detection of HC$_3$N emission toward the inner region of Mrk~231 raises interesting questions about regions of shielded gas near the nucleus of this ULIRG QSO. High-resolution observations will pinpoint the exact location of the HC$_3$N emitting regions, allowing us to better understand the underlying mechanisms and evolution of this powerful galaxy.


\section{Conclusions}

In this paper we present the first high-resolution observations of HCN, HCO$^+$ and HNC 1--0 emission from the wings of
the prominent molecular outflow of the QSO ULIRG Mrk~231. The velocities ($\approx$ 750 \kms) are similar to those already found
for the CO-outflow. The line wings are spatially extended by at least 0.\arcsec 75 ($>$700 pc) in a north-south direction. This
is the first time prominent HCN and HCO$^+$ emission has been detected in a large-scale galactic outflow. The east-west rotation
of the main disk is also evident in the HCN, HCO$^+$ and HNC 1--0 narrow component line emission.

Compared to CO 1--0, HCN 1--0 appears relatively enhanced in the line wings by factors of 2--5. This suggests that a large
fraction of the gas in the outflow is dense $n>10^4$ $\cmmd$, which is consistent with the molecular gas being compressed and fragmented by shocks. The HCN abundances may be significantly elevated ($X$(HCN)$>10^{-8}$) in the outflow, which can be caused by shocks and/or high temperature chemistry.

Alternatively, the excitation of HCN could be affected by mid-IR radiation, but we suggest that this is less likely because of 
the extended nature of the outflow. 

The HCN/HCO$^+$ line peak ratio is 1.3 for the line center and 2.2 in the line wings.  I(HCN)$>$I(HCO$^+$) is not uncommon
in outflowing gas in Galactic sources. The peak center HCN/HNC line ratio is 1.8 and
the line wing ratio is 2.8. This is within the range of what is typically observed in galaxies.

An unexpectedly bright HC$_3$N 10--9 line was detected in the central 400~pc of Mrk~231. This molecule is often associated
with young star-forming regions and tends to be destroyed by intense UV and particle radiation. We find it unlikely that the emission originates in the starburst region of Mrk~231 and instead suggest that the HC$_3$N emission
is emerging from a shielded, dust-enshrouded region within the inner 40-50~pc where the gas is heated to high temperatures (200 - 300 K) by the AGN.   
We speculate that this dusty, warm region represens the final stage of the obscured, X-ray absorbed accretion phase of the QSO
in Mrk~231.

\begin{acknowledgements}
      We thank the IRAM Director Pierre Cox for granting us DDT time for this project and the PdBI staff for excellent support. We furthermore thank the referee, Phil Maloney, for a careful and educating referee report that helped to improve the paper. Finally, we also thank Eduardo Gonz\'alez-Alfonso for fruitful discussions.
\end{acknowledgements}

\bibliographystyle{aa}
\bibliography{mrk231_ref}
\end{document}